\documentclass[twocolumn]{aastex631}

\newcommand{\msini}{{$m \sin i$}}

\usepackage{amsmath}
\usepackage{mathtools}
\submitjournal{ApJL}

\shorttitle{Cometary Enrichment of Exoplanets}
\shortauthors{Seligman et al.}
\begin{document}

\title{Inferring Late Stage Enrichment of Exoplanet Atmospheres from Observed Interstellar Comets}
 
\begin{abstract}

The discovery of the first two interstellar objects implies that, on average, every star contributes a substantial amount of material to the galactic population by ejecting such bodies from the host system. Since  scattering is a chaotic process, a comparable amount of material should be injected into the inner regions of each system that ejects comets.  For comets that are transported inwards and interact with planets, this Letter estimates the fraction of material that is accreted or outward-scattered as a function of planetary masses and orbital parameters. These calculations indicate that planets with escape velocities smaller than their current day orbital velocities will efficiently accrete comets. We estimate the accretion efficiency for members of the current census of extrasolar planets, and find that planetary populations including but not limited to hot and warm Jupiters, sub-Neptunes and super-Earths can efficiently capture incoming comets. This cometary enrichment may have important ramifications for post-formation atmospheric composition and chemistry. As a result, future detections and compositional measurements of interstellar comets will provide direct measurements  of material that potentially enriched a sub-population of the extrasolar planets. Finally, we  estimate the efficiency of this enrichment mechanism for extrasolar planets that will be observed with the \textit{James Webb Space Telescope} (JWST). With JWST currently operational and these observations imminently forthcoming, it is of critical importance to investigate how  enrichment from interstellar comet analogues may affect the interpretations of exoplanet atmospheric compositions.  
\end{abstract}

\author[0000-0002-0726-6480]{Darryl Z. Seligman}
\affiliation{Department of the Geophysical Sciences, University of Chicago, Chicago, IL 60637}

\correspondingauthor{Darryl Seligman}
\email{dzseligman@uchicago.edu}

\author[0000-0002-7733-4522]{Juliette Becker}
\affiliation{Division of Geological and Planetary Sciences, California Institute of Technology, Pasadena, CA 91125}

\author[0000-0002-8167-1767]{Fred~C.~Adams}
\affiliation{Physics Department, University of Michigan, Ann Arbor, MI 48109}
\affiliation{Astronomy Department, University of Michigan, Ann Arbor, MI 48109}

\author[0000-0002-9464-8101]{Adina~D.~Feinstein}
\altaffiliation{NSF Graduate Research Fellow}
\affiliation{Department of Astronomy and Astrophysics, University of Chicago, Chicago, IL 60637, USA}

\author[0000-0003-0638-3455]{Leslie~A.~Rogers}
\affiliation{Department of Astronomy and Astrophysics, University of Chicago, Chicago, IL 60637, USA}


\section{Introduction} \label{sec:intro}

Planetary enrichment from comet collisions is a process that has been extensively studied --- and even observed in real time  --- in the Solar System. In 1993, the comet Shoemaker-Levy 9 was discovered \citep{Shoemaker1993} and within a year it  collided with Jupiter \citep{Weaver1995}.  Observations of the Jovian atmosphere during and immediately following the collision revealed the presence of molecules that had never previously been detected there, including carbonyl sulphide and carbon monosulphide \citep{Lellouch1995,Noll1995}. More generally, previous work has estimated that the late heavy bombardment (LHB) would have enriched Jupiter and Saturn with mass increments of $0.15\pm0.04\,M_\oplus$ and $0.08\pm0.01\,M_\oplus$ from comets \citep{Matter2009}.

While the accretion of Shoemaker-Levy 9 was a historical astronomical event, it is not representative of the fate of typical solar system comets. The short period comets (SPCs), or ecliptic comets with low inclinations, originate in the Kuiper belt and are subsequently perturbed onto orbits interior to the orbit of Neptune \citep{Leonard1930,Edgeworth1943,Kuiper1951,Fernandex1980MNRAS,Duncan1988,Quinn1990}. They temporarily reside in the Centaur region on giant planet crossing orbits, until reaching the inner Solar System and becoming Jupiter Family Comets (JFCs) \citep{Hahn1990,Levison1997,Tiscareno2003,DiSisto2007,Bailey2009,DiSisto2009,Nesvoryn2017,Fernandez2018,SeligmanKratter2021}. Because the escape velocity of Jupiter is larger than its orbital velocity, close encounters with the giant planet typically scatter comets back into the Centaur region or into the inner Solar System, instead of resulting in a collision. In most models of the Solar System's formation, these scattering events with giant planets undergoing radial migration and/or instability ejected tens of earth masses worth of comets into interstellar space \citep{Hahn1999,Gomes2004,Tsiganis2005,Morbidelli2005,Levison2008,Raymond2020}.

The discovery of the first two interstellar objects, 1I/`Oumuamua and 2I/Borisov, confirmed that the ejection of planetesimals is a ubiquitous process. Although a variety of ejection mechanisms have been proposed,  their relative importance remains unclear. The net result is the production of a galactic population of interstellar objects with a spatial number density of $n_{o}\sim 0.1-0.2\,$\textsc{au}$^{-3}$ \citep{Trilling2017,Laughlin2017,Jewitt2017,Zwart2018,Do2018,moro2019a,moro2018,Levine2021}. One promising ejection mechanism is simple ejection via interactions with giant planets  \citep{Laughlin2017,Raymond2018b,zhang2020tidal}. Alternative scenarios include ejection from a stellar system exhibiting post-main-sequence evolution \citep{Hansen2017,Rafikov2018b} and from a circumbinary system \citep{Cuk2017,Jackson2017}. 

Because scattering is a chaotic process, the amount of material scattered inward should be roughly comparable to that ejected in systems that produce interstellar comets.  This balance implies that the population of interstellar comets is representative of the cometary bodies that are delivered from the exterior to the interior in nascent planetary systems.  As a result, extrasolar planets that are capable of efficiently accreting comets can potentially experience significant chemical enrichment and atmospheric erosion from this population \citep{Wyatt2020}. Since  compositional ratios and metallicities of planetary atmospheres are often believed to contain signatures of their formation locations \citep{Oberg2011}, this cometary enrichment mechanism may have non-negligible ramifications for interpreting observations of exoplanet atmospheres. 

The compositions of interstellar comets are thus representative of the material that potentially enriched extrasolar planets. For example, 2I/Borisov had more  CO than H$_2$O, and these two molecules were the dominant constituents of its volatile inventory \citep{Bodewits2020, Cordiner2020}. The composition of `Oumuamua was not measured  because no coma was detected \citep{Meech2017,Jewitt2017,trilling2018spitzer}. Based on the object's non-gravitational acceleration \citep{Micheli2018}, however, it is plausible that it was composed primarily of H$_2$ \citep{fuglistaler2018solid,seligman2020,Levine2021_h2}, N$_2$ \citep{jackson20211i,desch20211i}, or CO \citep{Seligman2021}.

This present work investigates the process of cometary enrichment of exoplanet atmospheres implied by the existence of interstellar comets. This Letter is organized as follows. In \S \ref{sec:collisions}, we calculate the efficiency with which a planet accretes comets as a function of its orbital and physical properties. In \S \ref{sec:exoplanets}, we identify currently known exoplanets that are susceptible to cometary enrichment through this mechanism. In \S \ref{sec:discussion}, we discuss the ramifications for upcoming  observations of exoplanetary atmospheres with JWST and summarize our findings. 

\section{Efficiency of Cometary Accretion}  
\label{sec:collisions}

 When comets from the exterior regions in a planetary system are transported inwards by gravitational perturbations, they will eventually cross the orbits of the planets. These comets can then  either be accreted by the planet through collisions or outward-scattered. In this section, we estimate the fraction of comets that will collide with a planet as a function of its orbit and the planetary and stellar properties. At the end of this section, we discuss the cases when comets persist in the inner system without being accreted or ejected. 
 
First, we clarify some terminology that  will be used in the remainder of this Letter. The term ``outward-scattering'' colloquially encompasses both (i) scattering that leads the comet to remain in the system with changed orbital parameters, and (ii) scattering that ejects the comet from the system. For the remainder of this Letter, we only use the term outward-scattering  to describe (ii). The initial events which perturb the comet onto a high eccentricity trajectory towards the interior of the planetary system is referred to as inward transportation.

We consider a planet with mass $M_{\rm P}$, radius $R_{\rm P}$, semi-major axis $a_{\rm P}$, and eccentricity $e_{\rm P}=0$. The comet is transported inward from a distant orbit, and  has semi-major axis $a_{\rm C}$ and eccentricity $e_{\rm C}$. The  cross section of the planet that will lead to accretion, $\sigma_{\rm A}$,  is given by 
\begin{equation}
    \sigma_{\rm A}=f_g \bigg(\pi R_{\rm P}^2\bigg) \,,
\end{equation}
where $f_g$ is  a gravitational focusing factor defined as
\begin{equation}\label{eq:focusing}
f_g=1+\bigg(\frac{2G M_{\rm P}}{R_{\rm P}   v_\infty^2}\bigg)\,.
\end{equation}
 The  relative speed of the comet when it crosses the orbit of the planet is proportional to the quantity $\sqrt{3-T}$, where $T$ is the Tisserand parameter  with respect to the planet  \citep{Carusi1990}. The Tisserand parameter with respect to a planet is approximately conserved during a close enocunter, and is given by
 \begin{equation}
     T=\frac{a_{\rm P}}{a_{\rm C}}+2\sqrt{\frac{a_{\rm C}}{a_{\rm P}}}\,\sqrt{1-e_{\rm C}^2}\,\cos(i_{\rm C})\,,
 \end{equation}
where $i_{\rm C}$ is the inclination of the comet. Since we consider comets with low $i_{\rm C}$, $\cos(i_{\rm C})\approx1$. Therefore, the asymptotic velocity $v_\infty$ in the frame of the giant planet is given by the expression
\begin{equation}\label{eq:velocity_comoving}
    v_\infty^2\big(a_{\rm C},e_{\rm C}\big)= \frac{G M_*}{a_{\rm P}}\bigg(3-\frac{a_{\rm P}}{a_{\rm C}}-2\sqrt{\frac{a_{\rm C}}{a_{\rm P}}}\,\sqrt{1-e_{\rm C}^2}\bigg)\,,
\end{equation}
where $M_*$ is the mass of the star. Comets with significant inclination have larger relative velocities and do not necessarily interact with the planet at perihelia. Moreover, $a_{\rm C}(1-e_{\rm C}^2)=p_{\rm C}(1+e_{\rm C})$, where $p_{\rm C}$ is the  periastron of the comet. Equation (\ref{eq:velocity_comoving}) relies on the assumption that the orbit of the planet and the comet  are in the same plane. This assumption is reasonable, because the ecliptic comets in the Solar System typically have low inclination. 

Since the comets are transported inward from the distant parts of the planetary system and the giant planets of interest reside at small to moderate semi-major axes, we assume that $a_{\rm P}\ll{a_{\rm C}}$. As a result, the eccentricity of the  orbit of the comets must be large (close to unity), so that they cross the  orbits of the planets on nearly radial trajectories.  Equation (\ref{eq:velocity_comoving}) can be approximated as
\begin{equation}
    v_\infty^2\big(a_{\rm C},e_{\rm C}\big)\simeq \Lambda\,\bigg(\frac{G M_*}{a_{\rm P}}\bigg)\,,
\end{equation}
where the coefficient $\Lambda$ is defined to be 
\begin{equation}\label{eq:lambda}
\Lambda=3- 2\,\sqrt{\frac{p_{\rm C}}{a_{\rm P}}}\,\sqrt{1+e_{\rm C}}\,.
\end{equation}
In the limiting case of high-eccentricity comets that barely intersect the orbit of the planet, where $e_{\rm C}\simeq 1$ and $p_{\rm C}=a_{\rm P}$, the coefficient $\Lambda\simeq 3-2\sqrt{2}$. Alternatively, if the comet would collide with the star on first approach and $p_{\rm C}\ll a_{\rm P}$, then $\Lambda\simeq3$. Therefore, for the comets considered in this Letter, $\Lambda\in(0.18,3)$. As a result, the gravitational focusing factor is given by 
\begin{equation}
    f_g\simeq  1 + \bigg(\frac{2M_{\rm P}\,a_{\rm P}}{\Lambda M_* R_{\rm P}} \bigg)\,
\end{equation}
and  the cross section for accretion has the form 
\begin{equation}\label{eq:sigma_a}
\sigma_{\rm A}=\pi R_{\rm P}^2\,
    \bigg(1 + \frac{2M_{\rm P}\,a_{\rm P}}{\Lambda M_* R_{\rm P}} \bigg)\,.
\end{equation}

In order to eject the comet out of the system, the escape velocity from the planet at the distance of closest approach $r$ must be greater than the escape velocity from the star system - which is related to the orbital velocity of the planet. This condition (which is necessary but not sufficient) can be written in the form 
\begin{equation}
    \frac{GM_{\rm P}}{r} \ge \frac{GM_*}{a_{\rm P}}\,.
\end{equation}
We define the quantity $R_{\rm S}$ to be the distance of closest approach to the planet that is required in order to scatter the object out of the planetary system. In other words, for outward-scattering to occur, we must have $r\le R_{\rm S}$, where 
\begin{equation}\label{eq:rs}
    R_{\rm S}=\Gamma \,\frac{M_{\rm P}a_{\rm P}}{M_*}\,,
\end{equation}
where $\Gamma$ is a dimensionless factor of order unity. An analogous result arises from the requirement that the impact parameter for outward-scattering must be comparable to the semi-major axis of the hyperbolic orbit of the comet around the planet (e.g., see the discussion of \citealt{Napier2021}). 

\begin{figure*}
\begin{center}
       \includegraphics[scale=0.6,angle=0]{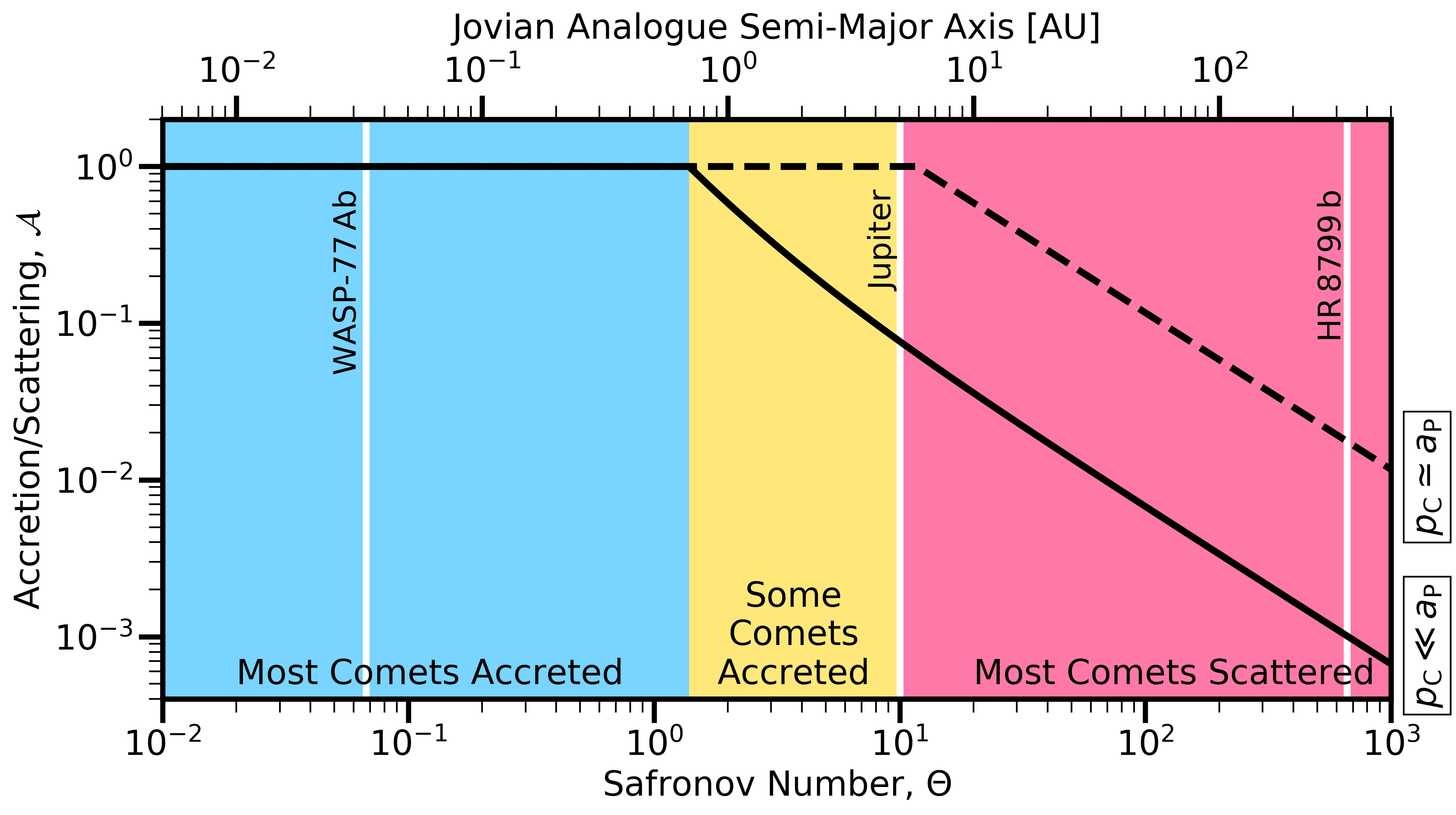}
    \caption{The ratio of accreted to outward-scattered comets, given by Equation (\ref{eq:fraction}), as a function of the Safronov number of the interacting planet. The (i) dashed and (ii) solid lines correspond to comets with perihelia (i) much less than ($\Lambda=3$, $\Gamma=1$)  and (ii) approximately equal to  ($\Lambda=0.18$, $\Gamma=1$) the semi-major axis of the scattering planet. The blue, yellow and red regions indicate where $\sigma_{\rm A}/\sigma_{\rm C}=100\%$, $10\%<\sigma_{\rm A}/\sigma_{\rm C}<100\%$ and $\sigma_{\rm A}/\sigma_{\rm C}<10\%$ for $\Lambda=3$. The top x-axis shows the semi-major axis corresponding to the Safronov number on the bottom axis, for a planet with a Jovian  mass and radius, orbiting a solar type star. The positions of WASP-77\,Ab, Jupiter and HR\,8799\,b are indicated by vertical white lines.}\label{fig:cross_section} 
\end{center}
\end{figure*}

In order to calculate the fraction of objects that collide, out of all the objects that interact with the planet, we compare the relative cross sections for accretion and outward-scattering. We denote the Safronov number by $\Theta$,  which is typically defined as
\begin{equation}
    \Theta= \frac{M_{\rm P}a_{\rm P}}{ M_*R_{\rm P}}\,.
\end{equation}
The two cross sections can then be written in the (approximate) forms 
\begin{equation}\label{eq:cross_sections}
    \begin{dcases}
    \sigma_{\rm A}\simeq\pi R_{\rm P}^2\,\big(1+2\Theta/\Lambda\big)\\
    \sigma_{\rm S}\simeq\pi R_{\rm P}^2\,\big(\Gamma^2 \Theta^2\big) \,.\\
    \end{dcases}     
\end{equation}
 
 The ratio of the number of comets accreted by the planet to the number comets that experience strong gravitational interactions with the planet (getting sufficiently close that they could be ejected from the system),  is given by the ratio of the cross sections, $\sigma_{\rm A}/\sigma_{\rm S}$. We denote this ratio as the accretion efficiency $\mathcal{A}$, which is given by
\begin{equation}\label{eq:fraction}
    \mathcal{A}=\min\bigg[1,\bigg(1+2\Theta\big/\Lambda\bigg)\,\bigg/\, \bigg(\Gamma^2 \Theta^2\bigg)\bigg]\,.
\end{equation}
When $\Theta \ll1$, $\sigma_{\rm A}/\sigma_{\rm S}$ given by Equation (\ref{eq:cross_sections})  can be much greater than unity. In this case, all comets that are sufficiently close to interact with the planet will be accreted.  We set $\mathcal{A}=1$ with the $\min$ function in Equation (\ref{eq:fraction}) in this case. 

This fraction only applies to comets  that have (i) a planet crossing orbit \textit{and}  (ii) a close encounter (within the outward-scattering cross section) at some point in time. This criteria guarantees that the comet experiences significant gravitational perturbations. However, the occurrence rate of these close encounters is low -- of order the ratio of the  diameter of the  maximum of the two cross sections  $\sigma_{\rm A}$ and $\sigma_{\rm S}$  to the orbital circumference --  even for a comet with a planet crossing orbit and low inclination. Nevertheless, the typical period of these cometary orbits, $P_{\rm C}$, will be approximately $P_{\rm C}\le 10^{2-3}$ years. For typically evolved planetary systems, the comets will thus orbit many times and will eventually experience close approaches with the planets. 

It is feasible that   the mechanism that initially transported comets into the interior of the system was significantly different than those that produce  ecliptic comets in the Solar System. In this case, some of the comets may have non-negligible inclinations,  making  accretion by a planet less common.  Specifically, when $\pi (\sin(i_{\rm C})\,p_{\rm C})^2>{\rm Max}[\sigma_{\rm A},\sigma_{\rm S}]$, the comet will typically not interact with the planet at perihelia. Similarly, we have  assumed an initial distribution of eccentricities for the scattered comets that eventually produce planet crossing orbits. However, the  distribution of eccentricities will also depend on the transport mechanism. Depending on the  mechanism, a significant  fraction of material could have  periastrons that are  too high to reach the outer-most planet under consideration and contribute to enrichment. It is also possible that any given close approach may \textit{not} result in ejection or accretion, and  the comet is simply perturbed outside of the cross sectional area. However, because of the differences in cometary orbit timescales and system ages, these comets are likely to (eventually) interact with a planet again.

A significant complication to this enrichment mechanism arises due to the finite lifetimes of the comets, commonly referred to as cometary fading. This process results primarily from  volatile mass loss, and less commonly from disintegration after multiple perihelia passages \citep{DiSisto2009,Brasser2015}. Typical mass loss rates from  active nuclear surface area are of order $\sim40 {\rm g\, cm^{-2}}$ per perihelia passage  of $\sim2$au \citep{Fernandez2005}. However, the active fraction of the surface  likely decreases with each perihelia passage \citep{Fernandez2005}. In the Solar System, once an object becomes a JFC\footnote{A common definition of a JFC is an object with Tisserand parameter with respect to Jupiter, $2<T_J<3$ \citep{Levison1996}. See Table 1 in \citet{SeligmanKratter2021} for an exhaustive list of literature definitions.}, the average dynamical lifetime is $\sim12,000$ years \citep{Levison1997}. Unfortunately, the efficiency of cometary fading remains unconstrained in (i) other planetary systems  and (ii) for different types of comets. Nevertheless, it is possible that a substantial fraction of the comets (or fragments) that enrich a planet are severely depleted in surface volatiles. However, subsurface   volatiles in cometary nuclei may remain intact indefinitely.

In Figure \ref{fig:cross_section}, we show the ratio of accreted to outward-scattered comets, given by Equation (\ref{eq:fraction}), as a function of the planetary Safronov number. The dashed and solid lines correspond to the limiting cases of $\Lambda=3-2\sqrt{2}$ and $\Lambda=3$, and $\Gamma=1$. For a given distribution of cometary eccentricities, the integrated accretion efficiency will be between these two curves. For both cases, this fraction falls quite steeply. As a leading order approximation, Jovian planets interior to $\sim1$au will experience enrichment from all comets that they interact with. When  $\Theta$ is large, the fraction of comets accreted is proportional to  $\Theta^{-1}\propto a^{-1}$, for fixed values of the planet and host star properties. This figure may be compared with Figures 1-4 in \citet{Wyatt2017}, who performed a similar analysis accounting for the more general cases of cometary accretion, ejection or retention for a range of stellar masses, planet masses and semi-major axes and reached similar conclusions.

For example, WASP-77\,Ab would have accreted all of the comets that it encountered, while the directly imaged massive planet HR\,8799\,b would have outward-scattered $\sim10^{2-3}$ times as many comets as it accreted. Interestingly, the atmospheric carbon to oxygen ratio of both of these planets has been measured to be 0.59$\pm0.08$ with a Solar value of 0.55 \citep{Line2021} and 0.578$\pm$0.005 \citep{Ruffio2021} with a Solar value of $0.54^{+0.12}_{-0.09}$ \citep{Wang2020}, respectively. However, \citet{Line2021} also measured a sub-solar overall metallicity in WASP-77\,Ab. This  suggests that the planet did \textit{not} experience significant cometary enrichment, which may be due to the lack of  external planets in the system.

Similarly, while the accretion efficiency for Jupiter is between $10-100\%$, \citet{Matter2009} calculated that the planet only accreted  $0.15\pm0.04\,M_\oplus$ of material during the LHB. In their simulations, the primordial  belt had $24\,M_\oplus$ of planetesimals when the instability that led to the LHB occurred. Our analytic calculation would predict Jupiter's accretion of $\sim1M_\oplus$ of planetesimals, which is inconsistent with the simulations by a factor of $\sim10$. This may be due to (i) outward-scattering events with Saturn, Neptune and Uranus and/or (ii) the initial distribution of eccentricities of the planetesimals inward-scattered by Neptune (see the discussion regarding eccentricity distribution in \S \ref{sec:discussion}). Both of these examples demonstrate that numerical calculations should be applied when investigating this enrichment mechanism for \textit{specific} planetary systems.

\section{Cometary Enrichment of Exoplanets}\label{sec:exoplanets}

\begin{figure}
\begin{center}
       \includegraphics[scale=0.45,angle=0]{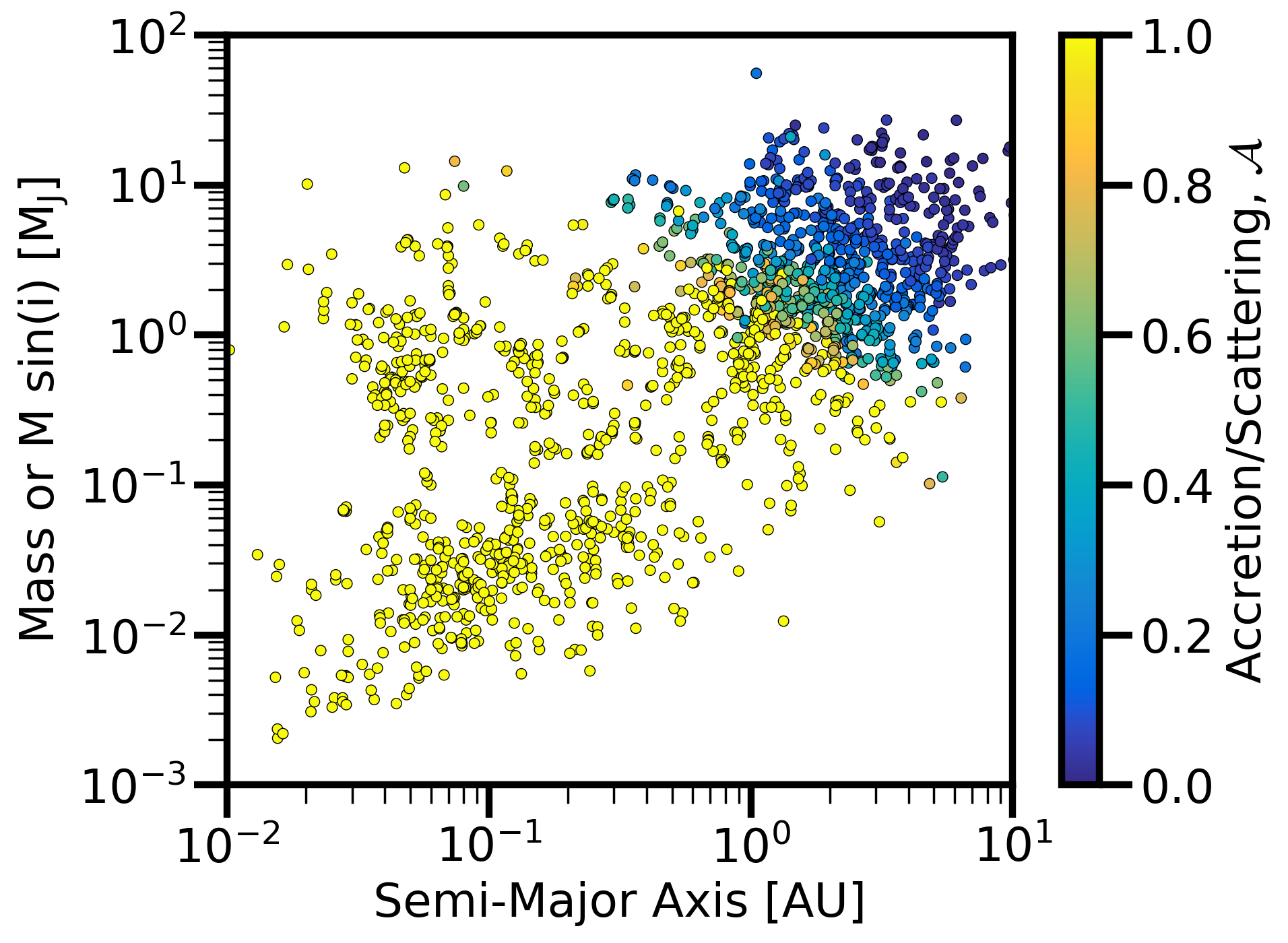}
\caption{The cometary accretion efficiency for every detected exoplanet. The x-axis shows the orbital semi-major axis, and the y-axis shows the mass or \msini\ for each planet. The color of each point indicates the efficiency with which the planet accretes comets, as given by the ratio of the accretion to outward-scattering cross section.  We calculate this quantity using Equation (\ref{eq:fraction}) with $\Lambda=1$ and $\Gamma=1$. }
\label{fig:exoplanets} 
\end{center}
\end{figure}

The detections of `Oumuamua and Borisov imply that, on average,  every star ejects approximately $M_{\rm C}\sim1-10M_\oplus$ of material into the galaxy \citep{Laughlin2017}. Since scattering is a highly chaotic process, a comparable amount of material should be scattered inwards as well, averaged across a large number of systems. 

In exoplanet-hosting systems, cometary material with orbits perturbed strongly enough by this process to be placed onto planet-crossing orbits will eventually pass within a planetary interaction cross section. The final fate of the comet (accretion or ejection) is approximately set by the ratio of $\sigma_{\rm A}/\sigma_{\rm S}$. Exoplanets with $\sigma_A/\sigma_S\sim1$  potentially experienced significant post formation envelope enrichment  from this population of comets.  In Figure \ref{fig:exoplanets}, we show this accretion efficiency for detected exoplanets. For planets that do not have  measured radii, we approximate it using the empirical mass radius relationship from \citet{Chen2017}. It is evident that hot and warm Jupiters, sub-Neptunes, and super-Earths are efficient at accreting  comets that they encounter. The population of Jovian mass planets with semi-major axes  $a>1$ au tend to have higher Safronov numbers, and  accrete a lower fraction of the cometary material that they experience close encounters with. 

Therefore,  compositions of interstellar comets may be representative of a fraction of the metal content of hot and warm Jupiters, super-Earths and sub-Neptunes. Assuming that the injected mass is  $\sim M_{\rm C}$ for every star, then multiplying this quantity by the fraction $\sigma_{\rm A}/\sigma_{\rm S}$ in Figure \ref{fig:exoplanets} will yield an \textit{upper limit} on the mass of cometary material from \textit{this population} that was accreted by any planet - assuming that every inward-scattered comet experienced a close encounter with the planet.  It is possible that additional cometary material was accreted, but this value of $ M_{\rm C}$ corresponds only to the material that is directly analogous to the galactic population of interstellar comets.

This result provides an upper limit, because our order of magnitude calculation does not take into account additional complications to the simplistic picture. In addition to the complications  discussed in \S\ref{sec:collisions}, one complication is that if there are multiple planets within a system, the comets will likely enrich the outer most planet that still has $\sigma_{\rm A}/\sigma_{\rm S}\sim1$. A second complication is that while $M_{\rm C}\sim1-10M_\oplus$ is estimated as an average value of material ejected per star, it is  likely that the   contributions from individual stellar systems vary.  Moreover, some fraction  (or even a significant fraction for some cases in Figure \ref{fig:cross_section}) of the inwardly transported comets are ejected in subsequent outward-scattering events. This effect can be reduced if inner planets have shorter scattering timescales \citep{Wyatt2017}, as confirmed with N-body simulations by \citet{Marino2018}. Generally speaking, tightly packed low mass planets were found to be more efficient at scattering material inward. These ejections would then also contribute to the interstellar comet population -- effectively reducing the initial flux of inwardly transported material needed to match that implied by 1I/`Oumuamua and 2I/Borisov. This effect would typically be less than a factor of $\sim2$, and within the uncertainties of this order of magnitude calculation.  This interpretation implicitly relies on the assumption that the majority of enrichment occurs \textit{after} the planet  reaches its currently observed location. This requires planetary migration to occur early. 


Atmospheric enrichment may also occur \textit{early} ($\tau\le10^7$ years) via planetesimal accretion in situ \citep{Zhou2007,Shibata2019,Podolak2020} or during migration \citep{Shibata2020,Shibata2022}. It may be possible to distinguish between planetesimal and comet accretion  based on the typical compositions of these objects. Their  compositions are a function of  formation location relative to various volatile condensation fronts \citep{Oberg2011,Seligman2022}. As an example, the atmospheric alkali metal content of hot Jupiters may indicate planetesimal accretion exterior to the H$_2$O snowline and subsequent inward migration \citep{Hands2022}. The population level compositions of interstellar comets are unconstrained, although 2I/Borisov exhibited a volatile C/O ratio indicative of formation exterior to the CO snowline \citep{Bodewits2020,Cordiner2020,Seligman2022}. If most interstellar comets form at large radial distances, interstellar cometary analogues and planetesimals  may deliver volatile material with varying condensation temperatures. However, the effects of cometary fading discussed earlier in this section may further complicate this  picture.


\section{Ramifications for Upcoming JWST Observations}\label{sec:discussion}

\begin{figure}
\begin{center}
       \includegraphics[scale=0.44,angle=0]{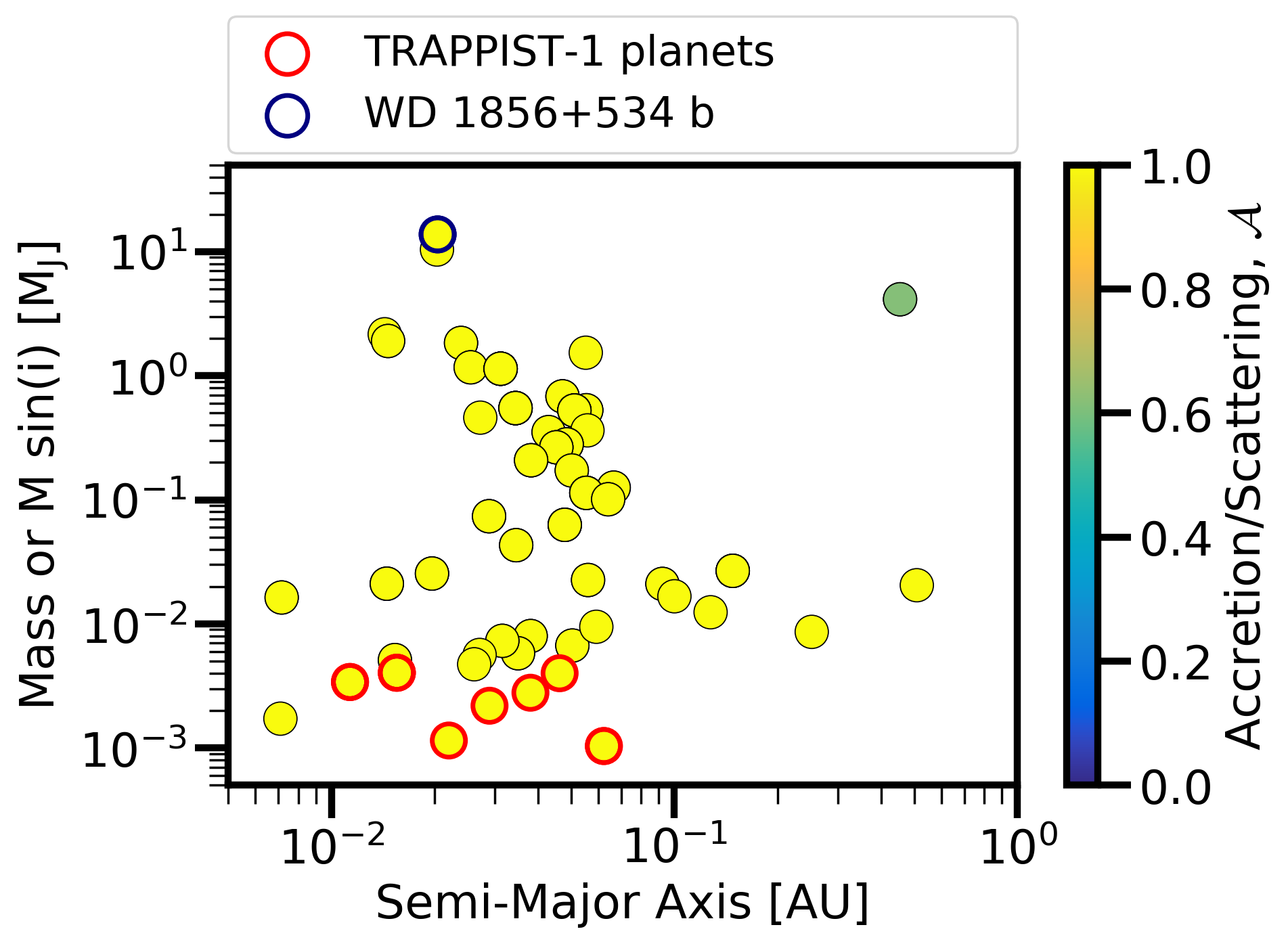}
\caption{The accretion efficiency for exoplanets that will be observed with JWST. Since most  targets are susceptible to cometary enrichment, this process should be considered when interpreting  forthcoming observations of these exoplanetary atmospheres. We calculate this  using Equation (\ref{eq:fraction}) with $\Lambda=1$ and $\Gamma=1$. Note that the following JWST targets are not included in the figure: TOI-836 b TOI-836 c, TOI-175 b, GJ 341 b, GJ 4102 b,TOI-134 b.}\label{fig:JWST TARGETS} 
\end{center}
\end{figure}

In this Letter, we showed that hot Jupiters, warm Jupiter, sub-Neptunes, and super-Earths  potentially experienced chemical enrichment from the accretion of a population of comets analogous to the interstellar comets. If every star, on average, ejected $M_{\rm C}\sim1-10M_\oplus$  of material in interstellar comets, then a comparable amount of material should also be transported inwards. Therefore, exoplanets that are capable of  efficient  accretion  may have experienced significant enrichment from these comets. 

This process could have important ramifications for  observations of exoplanet atmospheres, especially with  JWST  operational. For example, the atmospheric carbon-to-oxygen ratio is believed to trace the formation location of giant planets \citep{Oberg2011}. However, this ratio could  be significantly affected by cometary enrichment of up to $M_{\rm C}\sim10M_\oplus$ of metals. In Figure \ref{fig:JWST TARGETS} we show the same diagnostic of accretion efficiency as in Figure \ref{fig:exoplanets} for every planet that will be observed by JWST. It appears that almost every JWST target is  susceptible to significant cometary enrichment. We highlight the TRAPPIST-1 planets, which may have experienced both atmospheric destruction and subsequent replenishment from cometary accretion \citep{Kral2018}.


However, due to the non-exhaustive list of complications discussed  in the previous sections, the effects of this enrichment may be diminished for some or all of these planets.  Importantly, the  fraction of comets with planet crossing orbits, $\mathcal{F}$,  is determined by the eccentricity  distribution produced by the inward scattering process and $a_{\rm P}$ . A flat  eccentricity distribution results in $\mathcal{F}\simeq a_{\rm P}/a_{\rm C}\simeq1/40$. If the distribution is steeper or shallower, then $\mathcal{F}$ will be as high as $1/2$ or lower. Therefore, this effect could be diminished by $\sim1-50\%$ based on the eccentricity distribution of the comets. However, $\sim1\%$ accretion efficiencies could still enrich small planets with a mass of cometary material comparable to their own mass.  It is possible that by comparing  planetary and stellar metallicities,  we will be able to estimate the amount of cometary enrichment that a given planet experienced and therefore the extent to which the measured molecular abundances are primordial. 

The detection and compositional characterization of future interstellar comets will greatly refine the calculations and estimates presented in this Letter. The  carbon-to-oxygen ratio of the volatile coma of an interstellar comet can be used to infer its formation location within the protostellar disk relative to the CO snowline (Seligman et al. 2022). The fraction of interstellar comets that formed exterior and interior to the CO snowline will also represent the fraction of exoplanetary enriching comets that formed in these regions.

JWST will provide measurements of elemental molecular abundances and abundance ratios in these exoplanet atmospheres \citep[e.g. \textit{JWST} Cycle 1 approved programs][]{bean18, desert_jwst, hu_jwst, mann_jwst, jwst_mansfield, min_jwst, stolker_jwst}. With these observations scheduled within the next year, it will be of vital importance to quantify the effects of cometary enrichment on these exoplanets.  It is feasible that this enrichment could have provided a non-negligible fraction of the atmospheric metal content.

\facility {Exoplanet Archive}

\section*{Acknowledgements}

We thank Fred Ciesla, Megan Mansfield, Jacob Bean, Rafael Luque, Eliza Kempton, Samuel Cabot Dan Fabrycky, 
Sebasti\'{a}n Marino, 
Quentin Kral and Mark Wyatt for useful suggestions. We thank the  anonymous reviewer for extremely insightful comments and constructive suggestions that greatly strengthened the scientific content of this manuscript.

JB has been supported by the Heising-Simons \textit{51 Pegasi b} postdoctoral fellowship.
ADF acknowledges support by the National Science Foundation Graduate Research Fellowship Program under Grant No. (DGE-1746045). LAR gratefully acknowledges support from the Research Corporation for Science Advancement through a Cottrell Scholar Award.

This research has made use of the NASA Exoplanet Archive, which is operated by the California Institute of Technology, under contract with the National Aeronautics and Space Administration under the Exoplanet Exploration Program. This research also has made use of NASA’s Astrophysics Data System.

\bibliography{sample631}{}
\bibliographystyle{aasjournal}

\end{document}